# REPRODUCTION OF HYPOPNEA PHENOMENON USING A PHYSICAL AND NUMERICAL MODEL

F. Chouly[1], A.Van-Hirtum[2], P.Y.Lagrée[3], X. Pelorson[2] and Y. Payan[1]

ABSTRACT

Obstructive sleep apnea syndrome is now considered as a major health care topic. An in-vitro setup which reproduces and simplifies upper airway geometry has been the basis to study the fluid/walls interaction that leads to an apnea. It consists of a rigid pipe ("the pharynx") in contact with a deformable latex cylinder filled with water ("the tongue"). Air flows out of the rigid pipe and induces pressure forces on the cylinder. We present a numerical model of this setup: a finite element model of the latex cylinder is in interaction with a fluid model. Simulation of an hypopnea (partial collapsus of the airway) has been possible and in agreement with observations from the in-vitro setup. The same phenomenon has been simulated on a soft palate model obtained from a patient sagittal radiography. These first results encourage us to improve the model so as it could reproduce the complete apnea phenomenon, and be used for a planification purpose in sleep apnea surgery.

## 1. INTRODUCTION

Obstructive sleep apnea syndrome, which is due to intermittent closure of upper airway during sleep (Ayappa & al), has a serious impact on everyday life of more than 4% of men and 2% of women (Young & al), with consequences such as excessive daytime sleepiness and hypertension (Malhotra & al). Among therapeutic approaches for obstructive sleep apnea syndrome are surgeries such as uvulopalatopharyngoplasty (Fujita & al) or maxillo-mandibular surgery (Riley & al).

Success of a surgical procedure underlies on ability to forecast positive or negative effects of intervention on sleep apnea severity. Consequently, a mechanical model of the fluid (airflow) / structure (soft tissue) interaction which causes an apneic episode would be of great interest for surgical planning.

[1] TIMC Laboratory, UMR CNRS 5525, Université Joseph Fourier, 38706 La Tronche, France.
*{franz.chouly, yohan.payan}@imag.fr*
[2] Institut de la Communication Parlée, UMR CNRS Q5009, INPG, 38031 Grenoble Cedex, France.
*{annemie, pelorson}@icp.inpg.fr*
[3] Laboratoire de Modélisation en Mécanique, UMR CNRS 7607, B 162, Université Paris 6, 75252 Paris, France.
*pyl@ccr.jussieu.fr*

As a result, the aim of this preliminary work is to present a numerical simulation of hypopnea phenomenon, which is partial closure of upper airway with flow limitation (Gould & al). Results of these numerical simulations are qualitatively compared with observations on an in-vitro setup and from litterature.

2. MATERIAL AND METHODS

2.1. In-vitro setup

An in-vitro setup has been designed so as to reproduce soft-tissue deformation in response to fluid flow circulation in a constricted channel. It is made from a square rigid pipe and a latex cylinder filled with water (fig. 1 and fig. 2). Analogy might be established between the pipe and a pharynx, also between the cylinder and a tongue. Complete description of the setup can be found in (Van Hirtum & al). Alternatively, a rigid metallic cylinder can be adapted instead of the latex cylinder, so as to proceed to velocity and pressure measurements (Van Hirtum & al).

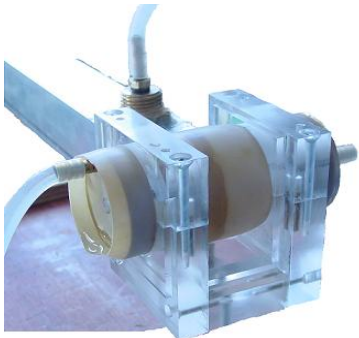

figure 1. In-vitro setup photography.

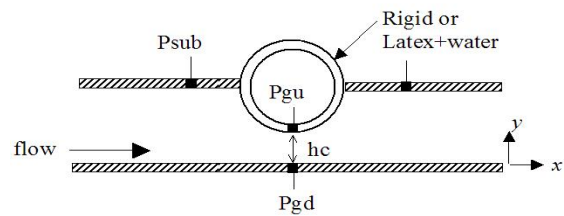

figure 2. In-vitro setup drawing. Psub is upstream pressure (measured), hc is height channel at constriction (controlled parameter), and Pgd is pressure at constriction (measured).

2.2. Sagittal radiography

From a pre-operative sagittal radiography of an apneic patient, a bidimensionnal approximation of the soft palate geometry has been obtained (fig. 4). This provides a more realistic geometry than the one in-vitro setup for simulations.

2.3. Structural models

First, a finite element model of the in-vitro setup has been designed (fig. 3). It reproduces geometrical and mechanical boundary conditions of the latex cylinder. A second model has been conceived from geometry of the soft palate (fig. 5). Boundary conditions have been chosen in agreement with anatomy (nodes constrained to immobility are those tighed to hard palate).

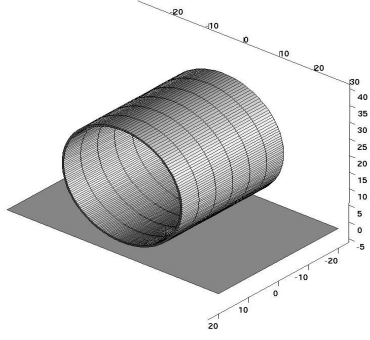 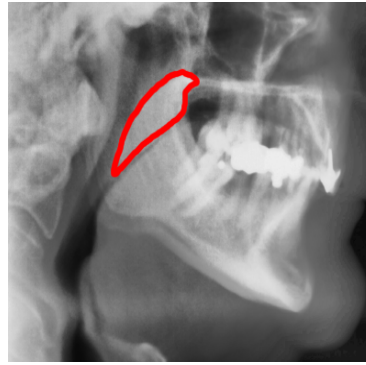 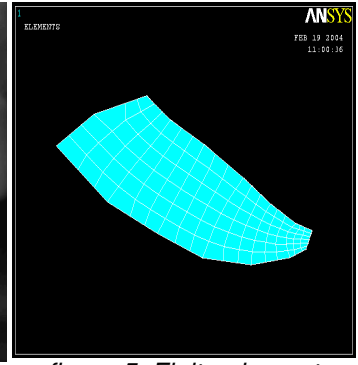

figure 3. Finite element model of the in-vitro setup (axes in mm).   figure 4. Patient sagittal radiography. In red: soft palate.   figure 5. Finite element model of the soft palate.

Both latex and soft tissue material have been considered as linear, homogeneous, isotropic, and nearly-incompressible. Young modulus are respectively $8.10^5$ Pa for latex (manufacturer data), and $5.10^4$ Pa for the soft palate (human soft tissue). Poisson's ratio is 0.499. Structural analysis, under small deformation assumption, has been carried out using Ansys(TM) Software.

2.4. Fluid models

Two fluid models have been confrontated for computation of pressure on the walls. The first is based on an analytical formulation:

$$p(x) = p_0 + \frac{1}{2}\rho \phi^2 \left( \frac{1}{A_0^2} - \frac{1}{A(x)^2} \right) \quad (1) \quad \phi \simeq c \cdot A_c \left( \frac{2}{\rho}(p_0 - p_s) \right)^{\frac{1}{2}} \quad (2)$$

where, for a given abscissa $x$ (fig. 2), $p(x)$ is pressure and $A(x)$ is channel section area. $p_0$ and $p_s$ are inlet and outlet pressure, $\rho$ is fluid density, $\phi$ is airflow rate, $A_0$ is initial section area, $A_c$ is constriction area and $c$ is a coefficient which determines position of separation of the fluid from the wall (Payan & al). It is obtained from Bernoulli equation as explained in (Payan & al) and in (Van Hirtum & al). The second model is based on a numerical resolution of the Reduced Navier-Stokes / Prandtl equations:

$$u\frac{\partial u}{\partial x} + v\frac{\partial u}{\partial y} = -\frac{\partial p}{\partial x} + \frac{\partial^2 u}{\partial y^2}$$
$$-\frac{\partial p}{\partial y} = 0 \quad (3)$$
$$\frac{\partial u}{\partial x} + \frac{\partial v}{\partial y} = 0$$

where $(u,v)$ are components of velocity vector, $p$ is pressure (non-dimensionnal equation with normalized variables). These equations are solved using finite difference method, as described in (Lagrée & al). This formulation is appropriate for boundary layer fluid flow circulation, which is the case in this context (Van Hirtum & al).

2.5. Interaction between structural and fluid models

An iterative algorithm enables fluid/structure interaction: pressure is computed using the fluid model. Deformation is computed from pressure forces with the structural model, which induces a new channel geometry for the fluid flow circulation, thus new pressure forces. The algorithm stops when no more significative deformation is computed (quasi-static assumption).

3. RESULTS

3.1. In-vitro setup

Deformations of the in-vitro setup latex cylinder (fig.6 and fig.7) in response to fluid flow circulation governed by Bernoulli equation for parameters given in table 1 have been simulated. Value of c = 1.05 has been chosen because it gives the best correlation between experimental data and model prediction on the in-vitro setup with a rigid cylinder (Van Hirtum & al). The channel at its narrowest part (constriction) becomes narrower (fig.8) and flow limitation can be observed from computation of air flow rate in response to pressure gradient (fig. 9). Same qualitative behaviour is obtained from simulations with RNS/P fluid model (fig. 8).

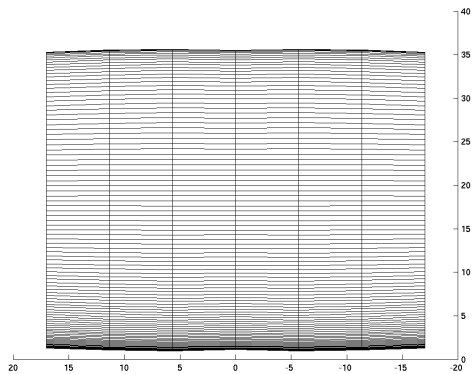

figure 6. Front view of the finite element model of the in-vitro setup at rest (axes in mm).

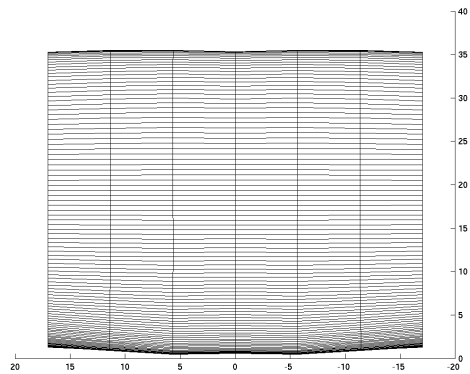

figure 7. Front view of the finite element model of the in-vitro setup after deformation (axes in mm).

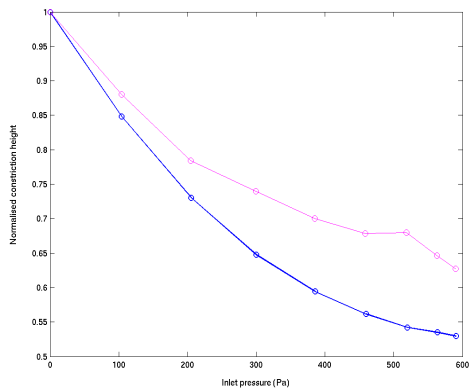

figure 8. Normalized constriction height for both Bernoulli (blue) and RNSP (magenta) fluid models.

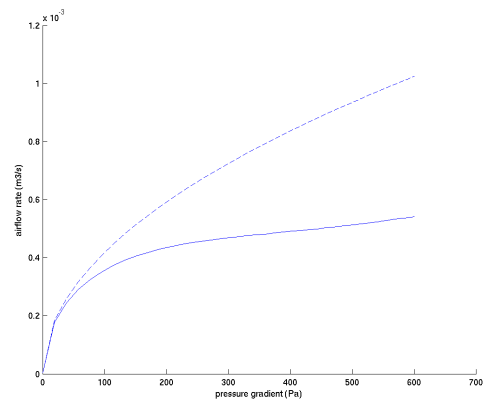

figure 9. Airflow rate / pressure gradient curves for both rigid (dashed line) and deformable (solid line) models.

3.2. Sagittal radiography

Simulations results of the soft palate model response to flow circulation for parameters given in table 1 are qualitatively the same as those of the in-vitro setup: decrease of the channel area at the narrowest site (fig. 10) and flow limitation. Value of separation coefficient (c = 1.2) has been taken from litterature (Payan & al).

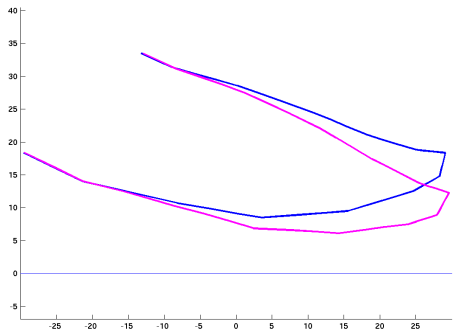

figure 10. Deformation of the velum in response to inspiration. In blue, soft palate at rest. In magenta, deformed soft palate (axes in mm).

| Parameter | Value (in-vitro setup) | Value (soft palate model) |
|---|---|---|
| pressure gradient | 600 Pa | 70 Pa |
| separation coefficient | 1,05 | 1,2 |
| constriction height | 1,1 mm | 8,5 mm |

table 1. Parameters for simulations.

4. DISCUSSION

First, both structural models adopt the same behaviour when in interaction with a fluid: decrease of the channel section at the level of constriction, from which results flow limitation. This corresponds to the hypopnea phenomenon as defined in (Gould & al). Moreover, for the in-vitro setup, good qualitative agreement has been observed between simulation results and reality.

Similar results are obtained from both Bernoulli and RNS/P fluid models although a significative quantitative difference has been observed in constriction height variation (fig. 8). Quantitative measurements of the latex cylinder deformation will be necessary to determine what model is the best at prediction. Neverthesless, RNS/P model seems more appropriate when looking at theoretical study and experimental measurements on rigid cylinder (Van Hirtum & al).

These encouraging results are the first step toward a more realistic model, which should integrate non-linearities of the structure and forces due to viscosity in the fluid (wall shear stress). Integrating contact forces might enable simulation of a complete apnea. Complete anatomy of upper airways obtained from CT-scans of apneic patients would be also helpful to proceed to simulation of hypopnea/apnea in more realistic conditions. Last but not least, ability of our modelling approach for quantitative prediction should be evaluated by comparing results of simulations and measurements on in-vitro setup with deformable latex cylinder.

AKNOWLEDGEMENTS

This work is partly funded by the Rhone-Alpes region, France.